\documentclass[pre,twocolumn]{revtex4}
\usepackage[dvips]{graphicx}
\usepackage{amsmath}
\usepackage{dcolumn}
\usepackage{bm}
\begin{document}
\title{Evidence for a Vanishing Complexity of the  Sherrington-Kirkpatrick model}
\date{\today}
\author{T.\ Plefka}
\affiliation{Department of Solid State Physics , TU Darmstadt, D
64289 Darmstadt, Germany}
\begin{abstract}
Based on the modified Thouless-Anderson-Palmer equations a
detailed numerical investigation for the complexity of the
Sherrington-Kirkpatrick  spin glass is worked out. The data
suggest a scaling law which leads to a vanishing of the complexity
in the thermodynamic limit as proposed recently. The results for
the total number of stable  states agree with existing approaches.
\end{abstract}
 \maketitle
An  important and common   feature  of all glassy systems with
quenched disorder is the presence of a huge number $ N_s $ of
stable or metastable thermodynamic states.  This number typically
grows exponentially
\begin{equation}\label{3}
N_s = \exp( N \Sigma^{tot})
\end{equation}
with the size of the system  $N$ and determines  the \textit{total
complexity} $ \Sigma^{tot}$. The number of states per free energy
$f$
\begin{equation}\label{1}
 N_s(f)= N_s \varrho_s (f) = \exp[ N \,\Sigma (f)]\,=\sum_{\alpha=1}^{N_s} \,\delta(f-
f_\alpha )
\end{equation}
defines both the normalized density $\varrho_s (f)$ and the
\textit{complexity}  $ \Sigma(f)$ where the  $f_\alpha$ are the
values of the free energy  per particle of the states labelled by
$ \alpha= 1,\ldots, N_s $. The complexity or the density
$\varrho_s (f)$ characterize the organization of the thermodynamic
states and  is in addition expected to be of fundamental
importance for an understanding of the dynamics in glasses
\cite{parisi03}.

The Sherrington-Kirkpatrick (SK) spin glass \cite{sk} is probably
the simplest model to  describe  the  physics of glasses.  This
model consists  of  $N$ Ising spins connected to each other by the
bonds $ J_{ij}$ which are independent random variables of variance
$ N^{-1}$ with zero means. Computation  of  the complexity for the
SK model  were published  more than two decades ago. The first was
one presented by Bray and Moore (BM) \cite{bm80}. Subsequent
approaches basically equivalent to BM are given in \cite{num}. In
these early approaches  a modulus sign of a Hessian determinant
was dropped without any serious justification.

 In a
series of papers \cite{roma,data,let} this procedure  has recently
been criticized claiming that inadequacies and even
inconsistencies result. Moreover such difficulties are also found
for an alternative, the so-called Becci-Rouet-Stora-Tyutin (BRST)
symmetric solution. Thus  Crisanti et al. (CLPR) \cite{let}
proposed that no complexity occurs in the SK model. Contrary  to
the latter scenario Bray and Moore \cite{bmc} and Aspelmeier et
al.\cite{abm} have very recently presented new arguments for the
validity of the BM approach  and concluded that this theory
remains a viable candidate  for the spin glass complexity. Thus
the analytical investigations of the complexity for the SK model
are at present extremely controversial.

In this letter  a numerical investigation of these problems is
presented with the aim to clarify this controversy. Recalling that
the thermodynamic states of the SK model are given by  the
solutions of the Thouless-Anderson-Palmer (TAP) equations
\cite{TAP}, the method is  obvious. Provided that the stable
solution of these TAP equations and their associated free energies
are known as function of the system size $N$ it is just  a simple
counting and an extrapolation to the thermodynamic limit $
N\rightarrow\infty$ which has to be performed to find  the
distribution $\varrho_s(f) $ and  the total number of states $
N_s$.

The main difficulty  is the explicit determination of the TAP
states. A solution of this problem has recently been proposed by
the present  author \cite{II}. This approach is  based on a
modification of the TAP equations which consistently describes
both  the stable solutions and the unstable solutions. Furthermore
this approach employs a fictive dynamics to find the TAP solutions
as fixed points of a set of equations of motion. These methods
have been used to calculate successfully various physical
quantities of the SK model \cite{III}. Even some hints for
behavior the complexity can be found in \cite{III}.

The latter methods are again used in the this work. Focusing
exclusively on  the complexity, the present investigation on this
subject is much  more detailed than \cite{III}. As the main result
of this letter, the numerical data  demonstrate that the density
$\varrho_s(f) $ tends to a $\delta$-function
\begin{equation}\label{2}
\varrho_s(f) \longrightarrow \delta(f -f_{eq}) .
\end{equation}
in the thermodynamic limit which in agreement with the CLPR
proposal implies that there is no complexity $\Sigma(f)$ in the SK
model. As a further result the numerical data  suggest  that the
\textit{total complexity} $\Sigma^{tot}$  \textit{exists and is
given by the BM theory}. Thus both controversial approaches, the
BM theory and CLPR proposal  agree in parts with the present work
although these approaches seem to be disjunct in the present form.

A vanishing complexity has interesting consequences. As all states
have the same  free energy value, none of them is
thermodynamically preferred. Such a system is   nothing else than
a \textit{multi-phase system}.  The phase (or the mixture of
phases) of any multi-phase system is generally  selected by the
(dynamic) history or by additional secondary interactions (like
conjugate or  symmetry breaking fields in conventional multi-phase
systems). Note that these mechanisms are characteristic for
glasses. Thus it should be possible to transfer at least some of
the well understood techniques for conventional multi-phase
systems to the SK spin glass.

After the presentation and the brief discussion of the main
results, some more details  are given in the remaining part of
this letter. With  reference to \cite{II,III} the essence of the
modified TAP approach and of the numerical procedure is reviewed
followed by  the analysis of the numerical data.

In zero external field the modified TAP equations for the local
magnetizations $m_i=\langle S_i\rangle_\beta\, $ are given by the
set of equations
\begin{equation}\label{4}
 m_i=\tanh \beta\big\{\sum_j J_{ij}m_j-m_i \chi_l\big\}
\end{equation}
 with the local susceptibility
\begin{equation}\label{5}
\chi_l =\frac{1}{N}\sum_i
\frac{\beta(1-m_i^2)}{1+\Gamma^2\,\beta^{2}(1-m_i^2)^2}.
\end{equation}
The quantity  $\Gamma$  is proportional to the density of zero
eigenvalues of the inverse susceptibility matrix and is determined
by
\begin{eqnarray}\label{6}
\Gamma &=& 0\quad \hspace{4.2cm}\mathrm{for}\,x\geq 0\\
1&=&\frac{1}{N}\sum_i
\frac{\beta^2(1-m_i^2)^2}{1+\Gamma^2\,\beta^{2}(1-m_i^2)^2}\qquad
 \mathrm{for}\,x\leq 0\label{7}
\end{eqnarray}
where
\begin{equation}\label{8}
x= 1-\beta^2( 1-2q_2+q_4)
\end{equation}
and where $ q_\nu=N^{-1}\sum_i m_i^\nu \quad (\nu=2,4) $ it
introduced. These equations are exact in the thermodynamic limit
but can approximatively be used for large finite $N$.

The condition $ x=0$ represents the central spin glass instability
condition \cite{at,bm,I}. Above the instability ($x\geq0 $) the
local susceptibility   reduces to the isothermal value
$\chi_l=\beta( 1-q_2)$ and the Eqs.(\ref{4})  are in complete
agreement with the original TAP equations \cite{TAP,I}. Essential
differences result below the instability $(x<0)$ as $\Gamma>0$ and
$\chi_l\neq\beta( 1-q_2)$   hold for the modified TAP equations.
This implies negative eigenvalues of the susceptibility matrix
\cite{II}. Thus all the states with $ x<0$ are unstable and have
therefore no relevance for thermodynamic quantities for
$N\rightarrow\infty$.

For the stable states the free energy per spin $ f(\{m_i\})$  is
given by the well known expressions \cite{TAP,I}
\begin{eqnarray}\label{82} &&f(\{m_i\})= \,-\,\frac{\beta}{2
N}\sum_{i\neq j}J_{ij}m_i
m_j\quad-\,\frac{\beta^2}{4} (1-q_2)^2\quad {}\quad\\
&+&N^{-1}\sum_i \Big\{\frac{1+m_i}{2}\ln
\frac{1+m_i}{2}+\frac{1-m_i}{2} \ln \frac{1-m_i}{2}\Big\}\nonumber
\end{eqnarray}
from which the free energy values $f_\alpha=f(\{m_i^\alpha\}) $ of
the TAP solutions $ \{m_i^\alpha\} \quad (\alpha=1,  \ldots ,N_s)
$ are obtained.

The explicit calculation of these  solutions $ \{m_i^\alpha\}  $
employs  the relaxational  Glauber dynamics
\begin{equation}\label{9}
 \dot{m}_i(t) = - m_i+ \tanh \beta \big \{\sum_j
J_{ij}m_j-m_i \chi_l (t)\big \}.
\end{equation}
where the local susceptibility $ \chi_l(t)$ is related to the $
m_i(t)$ via  Eq.(\ref{5}) and  Eq.(\ref{6})  or Eq.(\ref{7})
depending on the instantaneous value of $ x(t)$. The fixed points
solutions of Eqs.(\ref{9}) coincide with the TAP solutions and
both solutions exhibit  identical stability properties \cite{III}.
Note that the phenomenological Eqs.(\ref{9})  do not correctly
describe  dynamic effects (for microscopic equations of motion
compare \cite{DI,DII}). They are, however, simple and are
sufficient for the determination of the fixed point (or of  the
TAP) solutions.

The numerical procedure is analogous  to the former work
\cite{III} to which the reader is referred for details. Again the
integration routine `NDSolve' of Mathematica on workstations is
used to determine the  $ \{m_i^\alpha\} $ for a number of systems
with different sizes as listed in Tab.\ref{t1}. All runs start
with random initial conditions for the magnetizations $
\{m_i^\alpha\} $ and nearly exclusively binary distributions of
the bonds $ J_{ij}=\pm N^{-1/2} $ are used \cite{remark}. Although
data for other temperatures are available the results presented
  are restricted to  the temperature $T=0.2 $  following
the previous work \cite{roma,data,let}.

First the results for the  total complexity $\Sigma^{tot} $  are
given which are  (annealed) averages over  $N_{sys}$ systems. In
Tab.\ref{t1} the obtained values are listed  for sizes $N$ up to
150 spins. The errors resulting from the system to system
variations are found to be of the order of $10\%$ . For $N>150$
the numbers of runs are not sufficient to determine $\Sigma^{tot}
$ and at best some lower bounds can be given. From a strict point
of view all values for $\Sigma^{tot} $ are lower bounds as the
applied procedure does not guarantee that all existing solutions
are found. Based on investigations of some systems with a large
number of runs, however, these  errors can be estimated to be less
than the uncertainty  resulting from the system to system
variations. For these estimates  the decreases of the rate to find
 new solutions is analyzed  when one increases    the number of runs.
\begin{table}
\caption{\label{t1} Specifications and results: $ N$ and $
 N_{sys}$  are the values of the sizes and the
numbers of investigated systems respectively. The values for total
complexity $\Sigma^{tot}$ and the fraction $w^+$  for solutions
with $ x>0$ represent averages over the $ N_{sys}$ systems at the
temperature $T=0.2$.}
\begin{ruledtabular}
\begin{tabular}{rccrr}
$N $&$N_{sys}$&number of runs \footnotemark[1] &$\Sigma^{tot}\quad$&$w^+$(\%)\\
\hline 81& 98 &200 [25000] & $0.0545\quad$& 7.5
 \\
100& 57 &1000  [5000] & $0.0552\quad$ &7.5\\
120& 27 & 1000& $0.0512\quad$& 8.6 \\
150& 25 & 150 [6000]& $ 0.0509$ \footnotemark[2] &11.0\\
225& 10  & 250 [29000] & $ (0.0428)$ \footnotemark[3]&10.8\\
400& 6 & 250  [1000]& $ (0.0184)$ \footnotemark[3] & 9.0\\
600& 4 & 100 [400]&{\Large -}\hspace{5mm} &6.7\\
800& 2 & 120&{\Large -}\hspace{5mm} &7.7\\
\end{tabular}
\end{ruledtabular}
\footnotetext[1] {The values in bracket apply only to one or two
systems.} \footnotetext[2]{Average value calculated from 2
systems.} \footnotetext[3]{Lower bound for $\Sigma^{tot}$ found in
1 system.}
\end{table}

According to \cite{data}, the BM  value  and the BRST value for
the complexity at $T=0.2$ is given by  $\Sigma^{tot}_{BM}= .0522 $
and by $\Sigma^{tot}_{BRST}= 0.0021 $, respectively. Thus the
results for $\Sigma^{tot}$  presented in Tab.\ref{t1}  are in
agreement with the BM value but in disagreement with the BRST
value. The latter conclusion even applies to those cases where
only lower bounds are given.

\begin{figure}
\includegraphics[width=\linewidth]{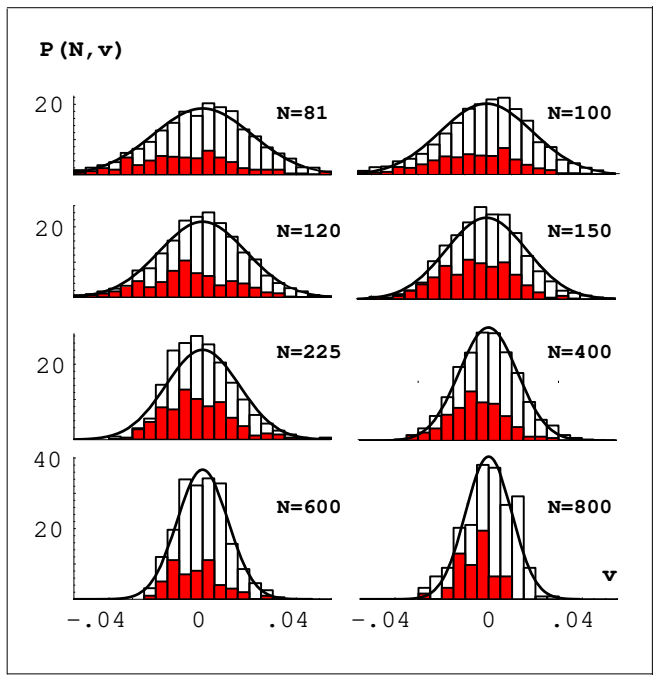}
\caption{\label{f1} Histograms corresponding to the distribution $
P (N,v)$ as defined in Eq.(\ref{10}) versus $v$ for various values
of the sizes $N$ at a temperature of $T=0.2$. The contributions
from solutions with $x>0$, magnified by a factor of four, are
indicated by the dark  areas. The curves are fits resulting from
the scaling law as described in the text.}
\end{figure}

\begin{figure}
\includegraphics[width=\linewidth]{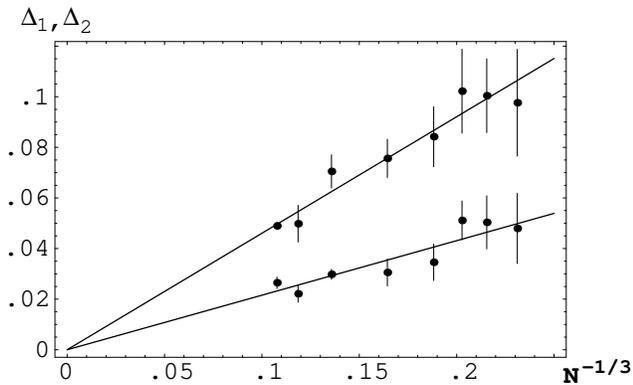}
\caption{\label{f2} Widths  $ \Delta_1= \langle
f_{max}^n-f_{min}^n\rangle$ (upper data) and $ \Delta_2= \langle
f^n_{min}-f_{av}^n\rangle$ (lower data) plotted versus $ N^{-1/
3}$ at a temperature $T=0.2$. The length of each error bar is two
standard deviations.}
\end{figure}

Next the histograms corresponding to the densities $ \varrho_s
(f)$ are analyzed.  It is noted that all the shapes of the
histograms for the individual systems (labelled by the index $n$)
look similar and are nearly Gaussian. Thus histograms of the
centered distributions
\begin{equation}\label{10}
P(v,N)=  \langle\varrho_s ^n(f_{av}^n +v)\rangle \quad
\textrm{with}\quad f_{av}^n=\frac{1}{N_{s}^n}  \sum_\alpha
f_\alpha
\end{equation}
averaged over the $N_{sys}$ systems
\begin{equation}\label{11}
\langle\ldots\rangle={N_{sys}^{-1}}  \sum_n \ldots
\end{equation}
are introduced \cite{rem0} and plotted in Fig.\ref{f1}. Without
any doubt the plots exhibit
 a narrowing of the $ P(v,N)$ with increasing values of $N$.
Moreover this narrowing is compatible with the scaling law
\begin{equation}\label{12}
N^{-s}\,P (N, N^{-s}v) =\tilde{N}^{-s}\, P(\tilde{N},
\tilde{N}^{-s}v)=P (1, v)
\end{equation}
for arbitrary values of $N$ and $\tilde{N}$. With a deviation of
less than one percent a  standard numerical fitting leads to the
value
\begin{equation}\label{13}
s={1}/{3}.
\end{equation}
Using for $P (1, v)$  a normalized Gaussian function with a
standard deviation of $0.09117$  the resulting fits are presented
in Fig.\ref{f1} showing a good agreement with the histograms.
Extrapolation of these results to $N\rightarrow\infty$ immediately
leads to the main result of Eq. (\ref{2}). Indeed the scaling law
(\ref{12}) implies $P(N,v)=N^s\;P(1,N^sv)$ which is a standard
representation of the $\delta$-function.

The scaling relation(\ref{12}) implies consequences for further
quantities. Denoting the extremal values of the free energy of a
specific system by $f^n_{min}$ and $f^n_{max}$  the  widths
defined by $ \Delta_1 = \langle f_{max}^n-f_{min}^n \rangle$ and
by $\Delta_2= \langle f^n_{min}-f_{av}^n\rangle $ should vanish
proportional to $N^{-s}$ in the thermodynamic limit. Indeed this
dependence is found according to Fig.\ref{f2}. Note that in this
analysis no assumptions on the shapes enter which further supports
the scaling relation.

\begin{figure}
\includegraphics[width=\linewidth]{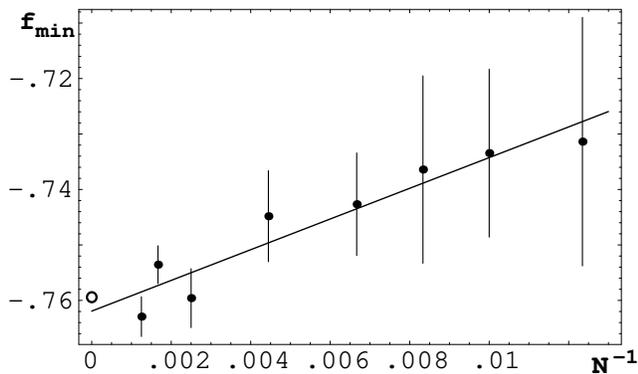}
\caption{\label{f3} Minimum value of free energy
$f_{min}=\langle\,f^n_{min}\rangle$   versus $ N^{-1}$ at a
temperature  $T=0.2$. The length of each error bar is two standard
deviations. The  equilibrium value of the replica breaking
approach is represented by the circle.}
\end{figure}

Fig.\ref{f3} shows  the minimum value of free energy
$f_{min}=\langle\,f^n_{min}\rangle$  averaged over the $ N_{sys}$
systems as function of $ N^{-1}$. In the thermodynamic limit the
width $ \Delta_2$ vanishes. Thus the limiting value of $f_{min}$
for $N\rightarrow\infty$ equals the value $f_{eq} $ which arises
in Eq.(\ref{2}). By linear extrapolation the numerical value  $
f_{eq}= -0.7619$ is obtained which is in good agreement with the
 equilibrium value $-0.7594 $ of the replica breaking approach
\cite{data}  at $T=0.2$.

Some points of the above analysis need  additional comments.
Analogous to the findings of \cite{III} the $ x$ values  of the
majority of solutions are negative. Indeed according to
Tab.\ref{t1} it is just a small fraction $w^+$ of the solutions
which satisfy the condition  $ x>0$. As discussed in \cite{III}
the solutions with $ x<0 $ arise due to finite size effects and
disappear with the power law $ |x|\sim N^{-2/3}$  for $
N\rightarrow\infty $ (compare Fig.2 of \cite{III}). This behavior
which also applies to the states with $x>0$ and  which is again
found in the present approach is nothing else but the well known
marginal stability of the TAP states in the thermodynamic limit.
As shown in Fig.\ref{f1} both the states with $x>0$ and the states
with $x<0$ are similar distributed. Thus a separate analysis of
the complexity  is not needed for these two types of states.

Recall that for  systems with large $N$ only a part of the TAP
states can be determined and that only this part enters in the
histograms of Fig.\ref{f1}. Therefore the interpretation of these
histograms as $\varrho_s(f) $ requires the assumption that this
part of the states  are representative  for the set of all states.
This assumption is checked for the systems where nearly all
solutions are found.  It is found that  the first hundred runs
give already reasonable approximations of the final results using
several thousands runs. No significant changes result for the
histograms, for the values $f_{av}^n$ and for $f^n_{min} $. Some
deviations are found at the upper tail of the distribution where a
few new solutions are obtained when increasing the number of runs
leading to  maximal changes of $20\%$ for $f^n_{max}-f^n_{min} $
and thus do not affect the basic results of this work.

In particular there is no hint that the range of the free energies
$ \Delta_1 $ increases to the BM value which is several times
larger than the typical numerical values. Moreover according to BM
results the number of these additional states should be large (a
finite fraction of the total number of $N_s$). Thus a finite
probability is expected to find at least some of these states
which however is not the case. These arguments also apply if one
takes in account that dynamical algorithms usually tend to prefer
the states with lower free energy \cite{ns}. This effect  is
clearly  observed in the present investigation. In disagreement to
\cite{abm}, however, it is concluded that this effect is not
sufficiently distinctive to explain the huge differences to the BM
approach.

Summing up, the numerical investigations  demonstrate that the
existing analytical approaches for the complexity of the SK model
are at best partially valid. The present work strongly supports
the CLPR proposal that there is no complexity for this model.
Finally the scaling relation suggested by the numerical results
may potentially be a guide for working out a consistent analytic
theory.

Interesting discussions with  A. Crisanti, B. Drossel, S. Kobe, L.
Leuzzi,  M. Moore and T. Rizzo are acknowledged.


\begin{references}
\bibitem{rem0}
This averaging is used to reduces the finite size effects.
According to the Figs.\ref{f2} and \ref{f3} the system to system
variations decrease with increasing $N$. This indicates the
expected self averaging in the thermodynamic limit.
\bibitem{remark} For $N=81$ a set of 28  systems with Gaussian distributions is
included for which the values of ground state energy are exactly
known \cite{kobe}. Using an extrapolation from $T=0.1 $ the
present algorithm  finds $82 \%$ of the exact values  within the
first 50 runs. This efficiency is remarkable as it is expected
that the ground state energy is missed in some cases due to saddle
node bifurcations \cite{III}.
\end{references}
\end{document}